\documentclass[aps,twocolumn,pra,superscriptaddress]{revtex4-2}
\usepackage{amsfonts}
\usepackage[final,hiresbb]{graphicx}
\usepackage{amsmath}
\usepackage{amssymb}
\usepackage{amstext}
\usepackage{color}
\usepackage{braket}
\usepackage{leftindex}

\usepackage{subfigure}
\usepackage{appendix}
\usepackage{bbm}
\usepackage{bbold}

\definecolor{Green}{RGB}{0,204,102}
\definecolor{Purple}{RGB}{102,0,255}
\definecolor{Blue}{RGB}{51,153,255}
\definecolor{Red}{RGB}{255,010,010}
\definecolor{Black}{RGB}{0,0,0}
\newcommand{\ml}[1]{{\textcolor{Black}{#1}}}

\begin{document}
	\title{Optical Polarization Holonomy in the Kerr Metric}
	
\author{Mark T. Lusk}
\email{mlusk@mines.edu}
\affiliation{Department of Physics, Colorado School of Mines, Golden, CO 80401, USA}

\begin{abstract}
Polarization holonomy is analytically determined for a class of closed, spherical trajectories of light transiting a black hole in the Kerr metric. The leading order geometric optics approximation admits a closed-form expression of such paths, and sets of source/receiver locations are quantified for a spectrum of black hole angular momenta. A conserved, conformal Yano-Killing scalar is then exploited to determine the evolving polarization. Polarization holonomy, the angle between outgoing and incoming polarizations, is quantified for the spectrum of admissible direct and retrograde trajectories. \ml{This offers a means of experimentally measuring Gravitational Faraday Rotation from a single, stationary position.}
\end{abstract}
\maketitle

	\section{Introduction}

The fabric of modern physics is woven from holonomies that characterize the structure of spacetime and parameter manifolds\cite{Wilczek1989, Berry1990, Nielsen2010}. These might be as conceptually immediate as the angle between initial and final orientations of a vector field parallel transported around a loop on a curved surface\cite{Oprea1995} or the geometric phase of a closed circuit on the Poincar\'{e} sphere\cite{Kwiat_1991}. However, they also manifest as Wilson loops for the non-Abelian gauge fields of the Standard Model\cite{Quigg2013}, underpin Loop Quantum Gravity\cite{Rovelli_1988}, and provide the requisite structure for Topological Quantum Computing\cite{Nayak2008}. 

\ml{Within the setting of stationary spacetimes, in particular rotating black holes, the polarization of light is predicted to exhibit} a helicity-dependent evolution that results in the rotation of linearly polarized planewaves. Such Gravitational Faraday Rotation has been \ml{theoretically} considered for decades\cite{Plebanski_1960, ThorneMisnerWheeler, FrolovShoom2011, Oancea_2020}, \ml{but the effect has yet to be experimentally measured. This is despite the investigation of extremely clever ideas that include quasar X-ray microlensing\cite{Chen_2015} and amplification of the effect using ring cavities\cite{Schneiter_2019}. Part of what makes measurements so difficult is that predictions are based on the transit of light between two observers\cite{FrolovShoom2011, Farooqui_2014}, and the requirement that a co-evolving reference frame be accounted for at either end.}  The consideration of closed trajectories would make this unnecessary, though, since the polarization can then be self-referenced. 

\ml{In particular, a projective holonomy can be quantified by comparing the outgoing and incoming polarizations at a common 3-space position of source and receiver. Circuits of this sort can be engineered from self-intersecting trajectories of constant radius. The ability to send and receive signals from a single point would allow precise measurements of the Gravitational Faraday Effect to be carried out for the first time, an important test of general relativity in the strong-field regime. Moreover, such polarization measurements can be made for light traveling on paths that range from nearly equatorial to those that boomerang from one pole to the other and back again. Both direct and retrograde orbits are possible, and they can even be made to pass through the ergosphere. A mapping of polarization rotation over a family of such trajectories amounts to the characterization of spacetime local to rotating black holes.} 

Recent work has made it straightforward to construct several morphologically distinct classes of trajectories using Jacobi elliptic functions\cite{Gralla_2020, Wang_2022}. One category is associated with non-trivial geodesics of constant radius, referred to as spherical trajectories. These self-intersecting paths can be cropped so that they start and end at the same point in 3-space, where the wave vector may be discontinuous. The initial and final polarizations may also be misaligned at such locations. 

For these closed circuits, the Walker-Penrose Theorem\cite{Walker_Penrose_1970, Penrose1973} can be used to analytically quantify polarization holonomy as a function of source position and black hole angular momentum. This allows holonomy to be quantified over the entire range of admissible direct and retrograde transits. The analytical approach and focus on holonomy distinguishes this work from that of earlier studies of light emanating from accretion disks\cite{Connors_1977} and the exchange of quantum information between a pair of observers near a Kerr black hole\cite{Farooqui_2014}. 

\section{Approach}
Because it can be exasperating to reconcile disparate notations, a brief review of the geometric optics approximation is provided along with a summary of the key scalars conserved in the Kerr metric due to isometries and hidden symmetries. Geometrized units are adopted to convert all physical quantities to length equivalents using black hole mass, M, speed of light, c, and gravitational constant, G\cite{ThorneMisnerWheeler}.  All such quantities are then non-dimensionalized using the mass length equivalent, $M G/c^2$. With black hole (Komar) angular momentum $J$, the singularity rotation rate is thus characterized by the non-dimensional parameter, $a = J c/(M^2 G)$\cite{Carroll2004}, which has a magnitude less than one in non-extremal settings\cite{Horowitz_2023}. 
  	
\subsection{Propagation of Light and Polarization within a Geometric Optics Approximation}

The source-free Maxwell equation with metric $g_{\mu\nu}$ is~\cite{ThorneMisnerWheeler}
\begin{equation}\label{MW4space1}
\nabla_\nu F_{\mu\nu} = 0 .
\end{equation}
Here $\nabla_\nu$ represents the components of the covariant gradient, while $F_{\mu\nu}$ is the skew-symmetric electromagnetic tensor, constrained by the Bianchi Identity\cite{Bianchi_1902} to be representable in terms of magnetic vector potential $A_\nu$: 
\begin{equation}\label{Bianchi}
F_{\mu\nu} = \nabla_\mu A_\nu - \nabla_\nu A_\mu.
\end{equation}
Applying the Lorenz gauge condition,
\begin{equation}\label{Lorenz}
\nabla \cdot {\bf A} = 0,
\end{equation}
and noting that the Ricci tensor, $R^\mu_{\,\,\nu}$, can be introduced using\cite{ThorneMisnerWheeler}
\begin{equation}\label{Ricci}
\nabla^\nu\nabla_\nu A^\mu = \nabla^\mu\nabla_\nu A^\nu + R^\mu_{\,\,\nu} A^\nu,
\end{equation}
we arrive at the de Rham form of the Maxwell equation:
\begin{equation}\label{deRham}
-\nabla^\nu\nabla_\nu A^\mu + R^\mu_{\,\,\nu} A^\nu =0.
\end{equation}

This representation lends itself to an Eikonal expansion of the vector potential, $\bf A$, in terms of the perturbation parameter $\epsilon:=\lambda/L$, the ratio of wavelength $\lambda$ and characteristic length $L$:
\begin{equation}\label{eikonal}
A_\mu = {\rm Re}\left[(a_\mu + \epsilon b_\mu + \epsilon^2 c_\mu+ ...)e^{i S/\epsilon}\right].
\end{equation}
The wave vector is defined as the phase gradient ${\bf p} :=\nabla S$. Coefficients $a_\mu$, $b_\mu$, etc. are functions of position, as is the phase accumulation $S$.
Substitution of this expansion into Eq. \ref{deRham}, and separation by powers of $\epsilon$, gives that the highest-order relation is independent of the de Rham tensor and is simply a statement that the wave vector has null character:
\begin{equation}\label{pnull}
{\bf p} \cdot {\bf p} = 0.
\end{equation}
Of course, this is equivalent to saying that the wave vector is parallel transported, 
\begin{equation}\label{pPX}
p^\nu \nabla_\nu p_\mu = 0,
\end{equation}
the geodesic equation that governs the trajectory of light within the lowest-order geometric optics approximation\cite{ThorneMisnerWheeler}.

Application of the Eikonal expansion, Eq. \ref{eikonal}, to the Lorenz condition of Eq. \ref{Lorenz}, implies that the polarization of light, ${\bf f} := {\bf a}/|{\bf a}|$, is orthogonal to the trajectory at leading order:
\begin{equation}\label{orthogpol}
{\bf f} \cdot {\bf p} = 0.
\end{equation}
Polarization is therefore parallel transported as well:
\begin{equation}\label{fPX}
p^\nu \nabla_\nu f_\mu = 0.
\end{equation}

The transport equations for wave vector and polarization each constitute a set of coupled, first-order ordinary differential equations (ODEs) that can be solved numerically.  While that is certainly one approach to determining a given trajectory and associated polarization, there is a better way to proceed. 

The evolution equation for trajectory can be re-expressed, using three conserved scalars, as four decoupled, first-order ODEs for spacetime position. These, in turn, can be solved in terms of Jacobi elliptic functions. The polarization along such paths can then be determined in closed form as well, using a \emph{hidden symmetry} associated with a conformal Killing-Yano tensor.  Once the Killing symmetries are explained, this is carried out in Kerr spacetime to engineer closed geodesics that exhibit polarization holonomy.

\subsection{Killing Fields}

\subsubsection{Isometries}

The strategy outlined above is made possible through the identification of operations for which the associated spacetime metric is invariant---i.e. isometries. This is equivalent to saying that the Lie derivatives of the metric are zero with respect to Killing vectors, $\bf v$, or Killing tensors, $\hat K$\cite{Carter1968, Walker_Penrose_1970, Chandrasekhar1998, Carroll2004, WiltshireVisser2009}:
\begin{equation}\label{Killing1}
{\cal L}_v g_{\mu \nu} = 0 , \quad  {\cal L}_K g_{\mu \nu} = 0.
\end{equation}
A more operationally useful form of these is
\begin{equation}\label{Killing2}
\nabla_{(\mu} v_{\nu)}  = 0  , \quad \nabla_{(\mu} K_{\nu \gamma)} = 0,
\end{equation}
where parentheses indicate the symmetry operator. Because the wave vector is also parallel transported, there is a conserved scalar, an integral of motion, associated with each such equation. 

In this work, attention is restricted to spacetime with a Kerr metric, expressed using oblate spheroidal Boyer-Lindquist coordinates\cite{BoyerLindquist1967} $\{t, r,\theta,\phi\}$ with covariant components of
\begin{equation}\label{KerrBL}
[g] = \left(
\begin{array}{cccc}
 \frac{2 r}{\Sigma }-1 & 0 & 0 & -\frac{2 a r  \mathbb{s}^2 }{\Sigma } \\
 0 & \frac{\Sigma }{\Delta } & 0 & 0 \\
 0 & 0 & \Sigma  & 0 \\
-\frac{2 a r  \mathbb{s}^2 }{\Sigma }  & 0 & 0 &  \frac{\mathbb{s}^2}{\Sigma} \left( (a^2+r^2)^2   - \mathbb{s}^2 \Delta a^2 \right) \\
\end{array}
\right).
\end{equation}
Here $ \mathbb{c} := \cos\theta, \quad \mathbb{s}  := \sin\theta$, while $\Sigma := r^2 + a^2 \mathbb{c}^2$ and $\Delta := a^2 - 2 r + r^2$ are standard in this setting.

Since neither time, $t$, nor azimuthal angle, $\phi$, appear in the Kerr metric, it has two Killing vectors, ${\bf v}_t$ and ${\bf v}_\phi$, with Boyer-Lindquist coordinates of $\{1,0,0,0\}$ and $\{0,0,0,1\}$, respectively. The associated conserved scalars are the spacetime counterparts to conserved scalars for energy and angular momentum:
\begin{equation}\label{Killing_Vectors}
\epsilon := -{\bf v}_t \cdot {\bf p}, \quad \ell := {\bf v}_\phi \cdot {\bf p}.
\end{equation}
%

\subsubsection{Hidden Symmetries}

There are also symmetries not associated with configurational isometries but, rather, symmetry operations in the dynamical state space. These are referred to as \emph{hidden symmetries}, and the associated operations are manifested in a generalization of Eqs. \ref{Killing2}, the conformal Killing-Yano equation\cite{Frolov_LRR_2017}:
\begin{equation}\label{KYeqn}
\nabla_{\mu} H_{\nu\lambda}  = \frac{1}{3}g_{\mu\nu}  \nabla^\gamma H_{\gamma\lambda}  -  \frac{1}{3}g_{\mu\lambda}  \nabla^\gamma H_{\gamma\nu} .
\end{equation}
The skew-symmetric 2-form, $\hat H$, is the \emph{Principal Tensor}\cite{Frolov_LRR_2017}, and its Hodge dual, $\hat F := \null^*\!\hat H$, is also a conformal Killing-Yano tensor. These fields are considered more fundamental than Killing tensors since their square always generates a symmetric Killing tensor,
\begin{equation}\label{KYprop}
K_{\mu\nu}  = F_{\mu\gamma} F_{\nu\alpha}  g^{\alpha\gamma},
\end{equation}
while the reverse is not necessarily true.

In Boyer-Lindquist coordinates, the covariant components are:
\begin{equation}\label{Htensor}
[H]=\left(
\begin{array}{cccc}
 0 & r & a^2 \mathbb{c} \mathbb{s} & 0 \\
 -r & 0 & 0 & a r \mathbb{s} ^2 \\
 -a^2 \mathbb{c} \mathbb{s} & 0 & 0 & a \mathbb{c} \mathbb{s} \left(a^2+r^2\right) \\
 0 & -a r \mathbb{s} ^2 & -a \mathbb{c} \mathbb{s} \left(a^2+r^2\right) & 0 \\
\end{array}
\right)
\end{equation}
and
\begin{equation}\label{Ftensor}
[F]=\left(
\begin{array}{cccc}
 0 & -a \mathbb{c} & a r \mathbb{s} & 0 \\
 a \mathbb{c} & 0 & 0 & -a^2 \mathbb{c} \mathbb{s} ^2 \\
 -a r \mathbb{s} & 0 & 0 & r \mathbb{s} \left(a^2+r^2\right) \\
 0 & a^2 \mathbb{c} \mathbb{s} ^2 & -r \mathbb{s} \left(a^2 + r^2\right) & 0 \\
\end{array}
\right).
\end{equation}
%

Eqs \ref{KYprop} and \ref{Ftensor} allow the Killing tensor components to be constructed as well:
\begin{equation}\label{Ktensor}
[K]=\left(
\begin{array}{cccc}
 a^2 \left(1-\frac{2 \mathbb{c} ^2 r}{\Sigma }\right) & 0 &
   0 & K_{t \phi} \\
 0 & -\frac{a^2 \mathbb{c} ^2 \Sigma }{\Delta } & 0 & 0 \\
 0 & 0 & r^2 \Sigma  & 0 \\
K_{t \phi} & 0 & 0 &
K_{\phi\phi} \\
\end{array}
\right),
\end{equation}
where
\begin{align}
&K_{t \phi} =   -\frac{a \mathbb{s} ^2 \left(a^2 \mathbb{c} ^2 \Delta +r^2
   \left(a^2+r^2\right)\right)}{\Sigma }\\
&K_{\phi\phi} =  \frac{\mathbb{s} ^2 \left(a^4 \Delta  \mathbb{c} ^2 \mathbb{s} ^2+r^2
   \left(a^2+r^2\right)^2\right)}{\Sigma } .
\end{align}
%
Finally, it is useful to construct a third conformal Killing-Yano 2-form, $\hat Z := \hat H + \imath \hat F$, with a matrix representation obtained immediately from Eqs. \ref{Htensor} and \ref{Ftensor}.

For any parallel propagated vector $\bf q$, tensors $\hat H$ and $\hat F$ exhibit scalar conservations of
\begin{equation}\label{ConservedScalars}
\bf q \cdot \hat H \cdot {\bf p} =0, \quad \bf q \cdot \hat F \cdot {\bf p} =0.
\end{equation}
More useful, though, is the conserved quantity, $\mathbb{k}$, associated with polarizations $\bf f$ and the Killing-Yano tensor $\hat Z$:
\begin{equation}\label{xi}
\mathbb{k}:={\bf f} \cdot \hat Z \cdot {\bf p} .
\end{equation}
%
This is the complex scalar of the Walker-Penrose Theorem\cite{Walker_Penrose_1970, Penrose1973}. Its magnitude is related to Carter's constant, $Q$,\cite{Carter1968, Chandrasekhar1998} by
\begin{equation}\label{Q}
Q = |\mathbb{k}| - (\ell - a\varepsilon)^2.
\end{equation}
For the Kerr metric in the Boyer-Linquist chart, $\mathbb{k}$ can be represented as\cite{Connors_1977, Chandrasekhar1998}:
\begin{equation}\label{k_Kerr}
\mathbb{k} = r \alpha -a \beta \mathbb{c}  -  \imath  ( r \beta + a \alpha  \mathbb{c} ),
\end{equation}
where
\begin{align}\label{k_Kerr_2}
&\alpha := f^1 p^0 + a (f^3 p^1 - f^1 p^3) \mathbb{s} ^2 \\
&\beta := -a f^2 p^0  \mathbb{s}  + (-f^3 p^2 + f^2 p^3) (a^2 + r^2)  \mathbb{s} .
\end{align}
%

\subsection{Trajectories of Light}

The most familiar application of Killing fields is associated with the construction of trajectories by identifying the wave vector as the derivative of position: ${\bf p} = d{\bf x}/dt$.  The Killing tensor, $\hat K$, satisfies Eq. \ref{Killing2}$_2$, and the conserved scalar is the magnitude of $\mathbb{k}$---i.e.
\begin{equation}\label{kmag}
 |\mathbb{k}| = {\bf p} \cdot \hat K \cdot {\bf p}.
\end{equation}
In Boyer-Lindquist coordinates, this is a first-order ODE in affine parameter $\tau$, for $dr/d\tau \equiv p^1$:
\begin{equation}\label{kmag2}
 |\mathbb{k} | = \frac{\left(\varepsilon  \left(a^2+r^2\right)-a \ell
   \right)^2}{\Delta }-\frac{(p^1)^2 \Sigma
   ^2}{\Delta }.
\end{equation}
%
%
Likewise, the Boyer-Lindquist expression for Eq. \ref{pnull} amounts to a first-order ODE in terms of both $p^1$ and $d\theta/d\tau \equiv p^2$:
\begin{align}\label{pnull2}
4 \varepsilon^2 r (a^2+r^2)&+2 a^2 \ell
   ^2 -8 a \varepsilon  r \ell +2 \Delta  \varepsilon
   ^2 \Sigma \nonumber \\
= \, \ell^2 ( a^2+\Delta )& \csc ^2\theta + 2
   \Sigma^2 ((p^1)^2 + (p^2)^2 \Delta ).
\end{align}
Two additional ODEs are supplied by the integrals of motion of Eq. \ref{Killing_Vectors} in terms of $dt/d\tau \equiv p^0$ and $d\phi/d\tau \equiv p^3$: 
\begin{align}\label{Killing_Vectors_2}
\varepsilon &= \frac{2 a p^3 r  \mathbb{s}^2}{\Sigma
   }-p^0\left(\frac{2 r}{\Sigma } - 1\right)  \nonumber \\
\ell &= \frac{ \mathbb{s}^2 \left(p^3
   \left(\left(a^2+r^2\right)^2 - a^2 \Delta  \mathbb{s}^2\right)-2 a p^0 r\right)}{\Sigma
   }.
  \end{align}

Eqs. \ref{kmag2}, \ref{pnull2},  and \ref{Killing_Vectors_2}$_{a,b}$, comprise four first-order ordinary differential equations for the four trajectory components\cite{StewartandWalker1973}. While they can be written as separate ODEs for each position coordinate, the equations are still coupled because $\Sigma = r^2 + a^2 \mathbb{c} $. This can be removed with a simple re-scaling of parametrization from $\tau$ to the Mino parameter, $s$\cite{Mino_2003}:
\begin{equation}\label{Mino}
\frac{dx^\mu}{ds} := \frac{\Sigma}{\epsilon} p^\mu \equiv \frac{\Sigma}{\epsilon}  \frac{dx^\mu}{d\tau} .
\end{equation}

It is also useful to observe that there are only two independent scalars in these equations, impact parameter $\lambda := \ell / \varepsilon$ and Carter ratio $\eta:= Q / \varepsilon^2$. The resulting equations of motion are
\begin{align}\label{EoM}
\left(\frac{dr}{ds}\right)^2 &= {\mathcal R}(r) \\
\left(\frac{d\theta}{ds}\right)^2 &= \Theta(\theta) \\
\frac{d\phi}{ds} &= \frac{a}{\Delta}(r^2+a^2 - a \lambda) + \frac{\lambda}{\mathbb{s}^2} - a \\
\frac{d t}{ds} &= \frac{(r^2+a^2)}{\Delta}(r^2+a^2 - a \lambda) + a(\lambda - a \mathbb{s}^2),
\end{align}
where
\begin{align}
{\mathcal R}(r) &=  \left(a^2-a \lambda
   +r^2\right)^2-\Delta 
   \left((\lambda -a)^2+\eta
   \right)\\
\Theta(\theta) &= \eta - (a \mathbb{c})^2 - \left( \lambda \mathbb{c}/\mathbb{s}\right)^2 .
\end{align}
Given a position for the light source, and characterizing its direction with parameters $\lambda$ and $\eta$, the first two equations can be solved independently to obtain the radial and polar trajectories. The results can then be substituted into the second pair to determine the azimuthal and temporal trajectories.

\subsection{\ml{Evolution of Polarization}}

The equations of motion and their solution are agnostic with regard to evolving polarization. The utility of the complex-valued Fermi-Walker constant of Eq. \ref{k_Kerr}, $\mathbb{k}$, is that it allows the polarization of light to be determined along any such path.  Since it is complex valued, its real and imaginary components provide two equations for the four components of the polarization vector, $\bf f$, provided that the initial polarization is known. Specifically, Eq. \ref{k_Kerr} allows us to construct functions for the real and imaginary parts of $\mathbb{k}$:
\begin{equation}\label{k_KerrRI}
{\mathbb k}_R({\bf f}, {\bf p}, \theta) = r \alpha -a \beta \mathbb{c} , \quad
{\mathbb k}_I({\bf f}, {\bf p}, \theta)  = -( r \beta + a \alpha  \mathbb{c} ).
\end{equation}
The terms $\alpha$ and $\beta$, defined in Eq. \ref{k_Kerr_2}, are functions of polar angle, $\theta$, polarization, $\bf f$, and the unit tangent to the trajectory, $\bf p$. 
\begin{align}\label{eqs1and2}
{\mathbb k}_R({\bf f}_i, {\bf p}_i, \theta_i) &= {\mathbb k}_R({\bf f}(s), {\bf p}(s), \theta(s) ) \nonumber \\
{\mathbb k}_I({\bf f}_i, {\bf p}_i, \theta_i)  &= {\mathbb k}_I({\bf f}(s), {\bf p},  \theta(s) ).
\end{align}

A third equation is generated by the orthogonality of polarization and the wave vector---i.e. $\bf f \cdot \bf p = 0$: 
\begin{equation}\label{eq3}
f_3 = \frac{\rho ^4 \left(\Delta  f_2 p_2+f_1
   p_1\right)}{\mathbb{s}^2 \Delta \left(2
   a p_0 r-p_3
   \left(\left(a^2+r^2\right)^2-a^2 \mathbb{s}^2 \Delta \right)\right)}.
\end{equation}
Here $\rho^2 = r^2 + a^2 \mathbb{c}^2$ and $\Delta = r^2 - 2 r + a^2$.

Finally, tangents to the trajectory are null vectors, so the polarization is only unique modulo a factor of the wave vector. The factor can always be chosen so that the temporal component of the polarization is equal to zero:
\begin{equation}\label{eq4}
f_{0,i} = f_{0}(s) = 0
\end{equation}
Given a trajectory and initial polarization, $f_3$ and $f_0$ of Eqs. \ref{eq3} and \ref{eq4} can be substituted into Eqs. \ref{eqs1and2} which are subsequently solved for $f_1$ and $f_2$ as a function of path position, $s$, on the trajectory. This is a primary application of the Walker-Penrose theorem. The final pair of equations are linear in $f_1$ and $f_2$, but the analytical expressions obtained are unwieldy and lacking in immediate physical insight on their own. Since they are easily obtained with symbolic algebra software, the explicit expressions are not written here. They are used, though, in the application to follow.

\section{Application}

\subsection{Closed Spherical Trajectories}

As detailed in a recent study, algebraic expressions for trajectories of constant radius can be constructed using Jacobi elliptic functions\cite{Wang_2022}. This reduces the number of 3-space components to two, making it easier to identify closed circuits that exhibit polarization holonomy. For such transits, the impact parameter $\lambda$ and Carter ratio $\eta$ are
\begin{align}\label{isorad_1}
& \lambda \equiv \frac{\ell}{\varepsilon} := a + \frac{r}{a}\biggl(r - \frac{2\Delta}{r-1}\biggr) \\
& \eta \equiv \frac{Q}{\varepsilon^2} := \frac{r^3}{a^2}\biggl(\frac{4\Delta}{(r-1)^2} - r\biggr) .
\end{align}
This implies that a path is completely fixed once an initial position and rotation parameter, $a$, are prescribed. The trajectories are given by the following equations\cite{Gralla_2020, Wang_2022}, with requisite functions defined in the Appendix:
\begin{align}\label{traj1}
\theta(s) &= \cos^{-1} \left[ -\nu_\theta \sqrt{u_+} \,\,{\rm sn} \left(  \sqrt{-a^2 u_-} (s + \nu_\theta \mathcal{G}_\theta ), \frac{u_+}{u_-} \right) \right] \nonumber \\
\phi(s) &= \lambda G_\phi  \nonumber\\
 + &\frac{2 a}{r_+ - r_-} \left[ \left( r_+ - \frac{a \lambda}{2} \right) \frac{s}{r - r_+} - \left( r_- - \frac{a \lambda}{2} \right) \frac{s}{r - r_-} \right] \nonumber \\
t(s) &= I(s) + a^2 G(s) . 
\end{align}

\begin{figure}[t]
	\begin{center}
		\includegraphics[width=0.75\linewidth]{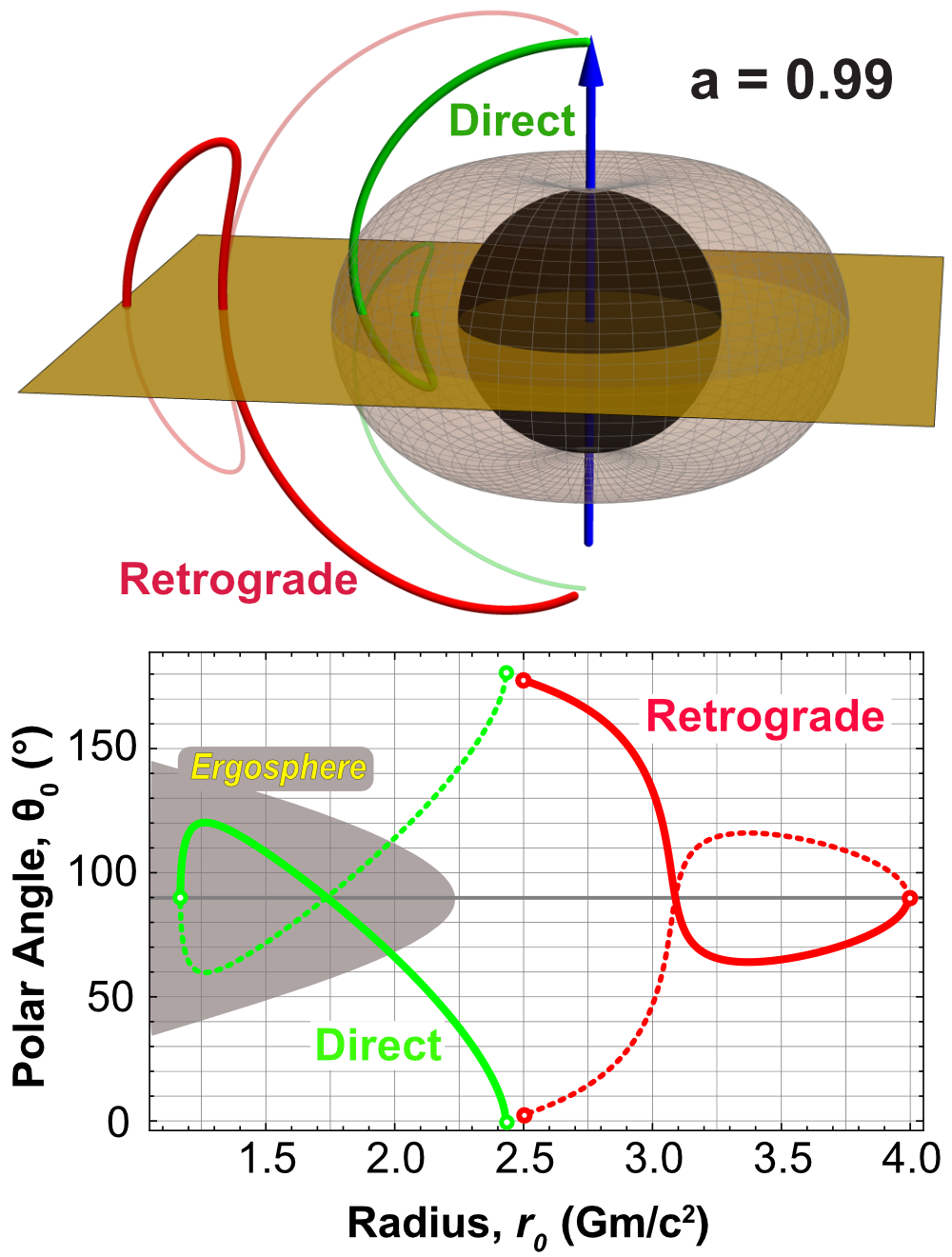}
	\end{center}
	\caption{ \emph{Send/receive points resulting in closed circuits.} Sets of retrograde (red) and direct (green) start/end points are shown for $a = 0.99$. Green and red curves that are semi-transparent (top) and dashed (bottom) identify admissible complementary sets ($\theta\rightarrow 180^\circ - \theta$). The top plot gives a 3-space projection of the positions, with the outer ergosphere shown in gray and the outer event horizon in black. Note that, for the direct case, it is possible to start/end within the ergosphere even though the light may escape it before returning.} 
	\label{start_end_set_a0p99}
\end{figure}

The travel of light, on a path of constant radius, will eventually overlap with itself.  This feature can be exploited to engineer a class of trajectories for which each member starts and ends at a specific radius and polar angle. Suppose that these and the rotation parameter are specified. Set the initial azimuthal angle to zero, and solve Eq. \ref{traj1}$_2$ for the value of Mino parameter, $s$, at which $\phi = 2 \pi$. For a given value of $a$, the points for which this is possible comprise an \emph{initial position arc} in spacetime. For instance, the set of all send/receive points with $a = 0.99$ is shown in Fig. \ref{start_end_set_a0p99}.

%
%

\begin{figure}[t]
	\begin{center}
		\includegraphics[width=1.0\linewidth]{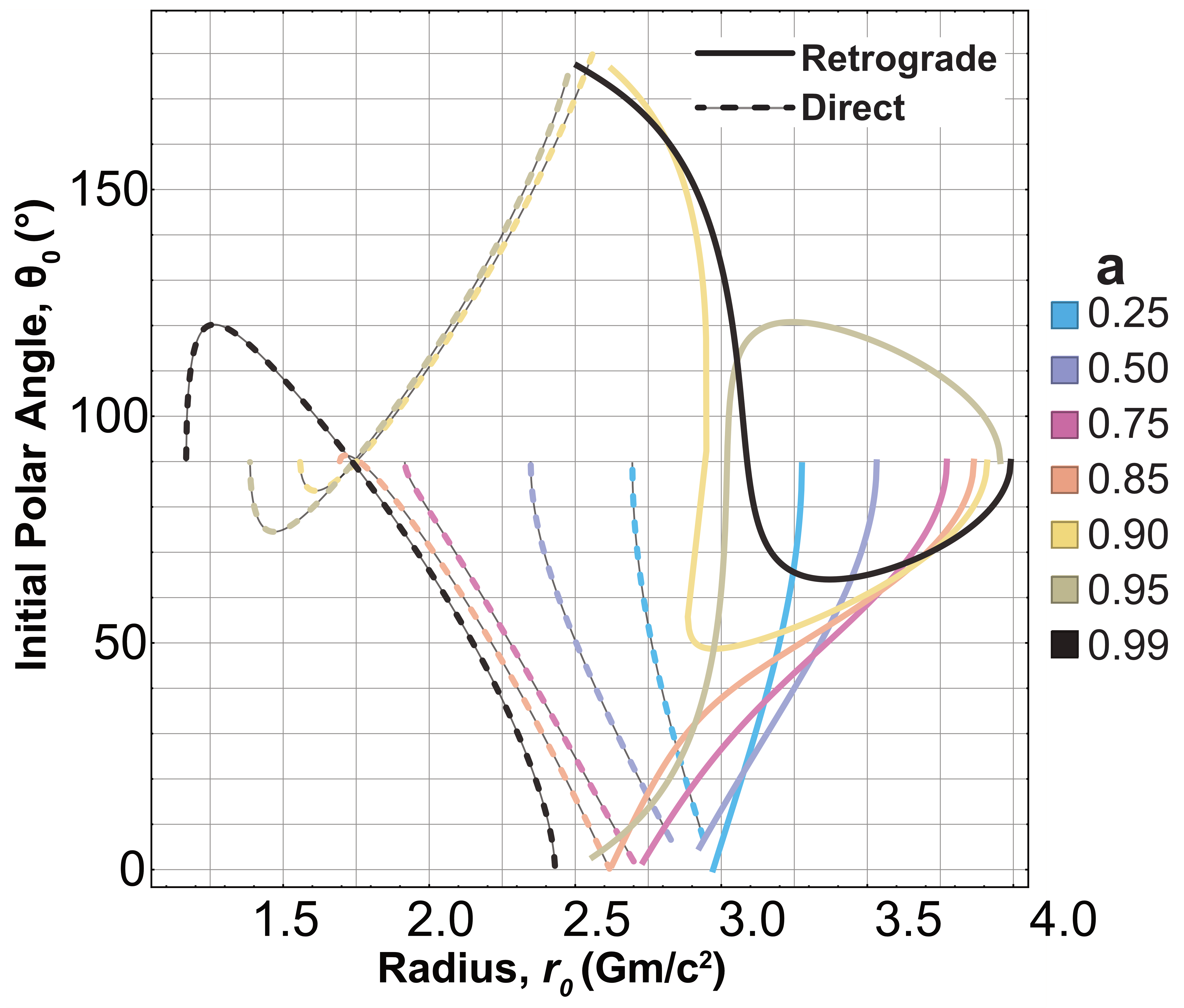}
	\end{center}
	\caption{ \emph{Source/Observation Points}. Admissible initial/final states are plotted for several values of rotation parameter $a$. Positions resulting in closed retrograde trajectories are shown with solid curves while positions that produce closed direct trajectories are dashed with a light gray outline to guide the eye. Complementary sets ($\theta\rightarrow 180^\circ - \theta$) are not plotted for the sake of clarity.} 
	\label{Initial_States}
\end{figure}

The methodology was then applied to construct initial position arcs for direct and retrograde paths over a range of black hole rotation values. These are shown in Fig. \ref{Initial_States}, where complementary sets ($\theta\rightarrow 180^\circ - \theta$) are not plotted for the sake of clarity. Notice that the range of admissible initial radii decreases with decreasing rotation rate, with an asymptote at the Schwarzschild black hole radius for unstable circular orbits $r = 3$.

%
%
%

\begin{figure}[t]
	\begin{center}
		\includegraphics[width=0.86\linewidth]{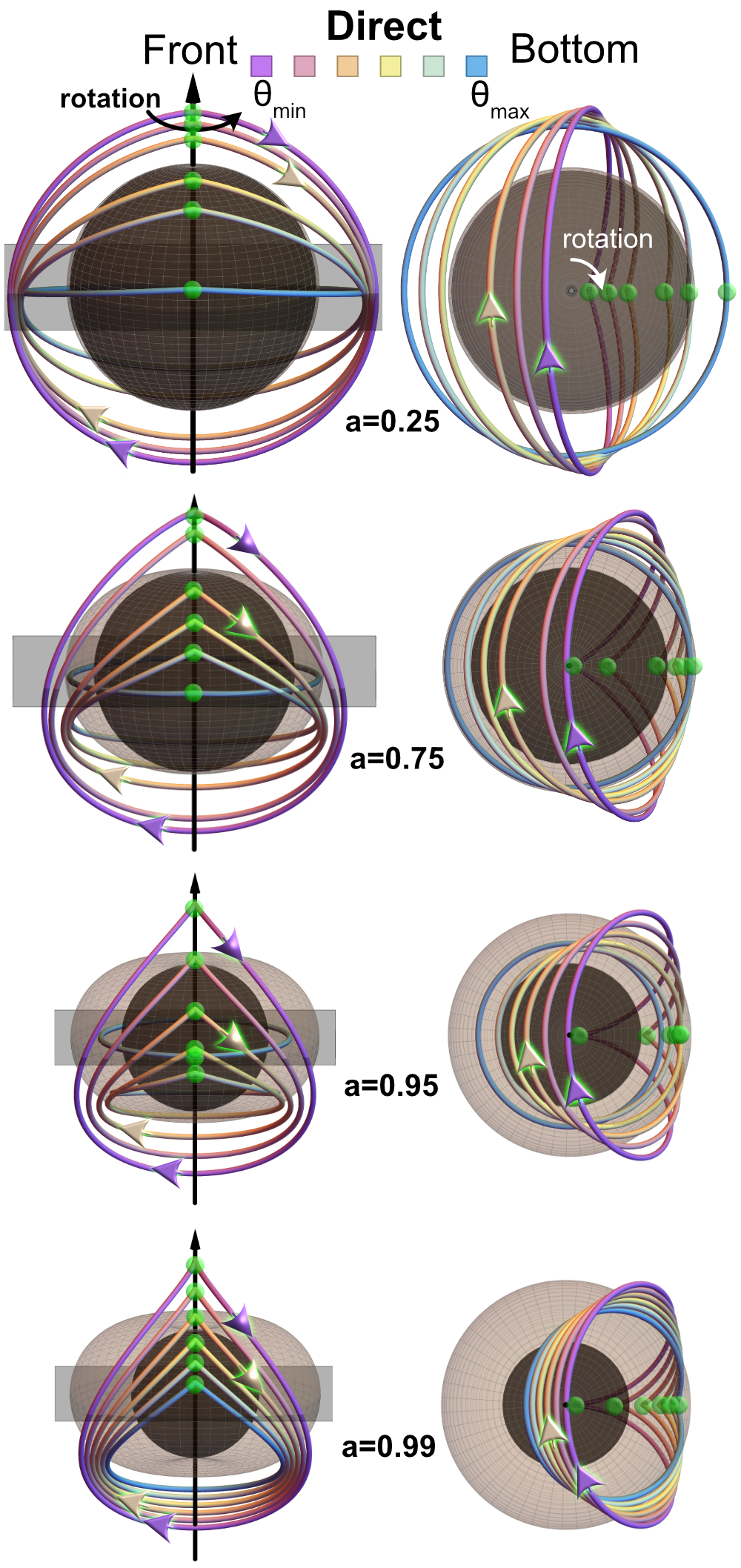}
	\end{center}
	\caption{ \emph{3-space projection of \underline{direct} trajectories.} Views are from front (left) and below (right). Arrows highlighted in green indicate direction of propagation, while the direction of black hole rotation is identified in the top panel of each column. The starting point of each trajectory is highlighted with a green sphere. Each row is associated with a particular value of rotation parameter, $a$. Panels include the outer ergosphere (light gray) and the outer event horizon (dark gray).} 
	\label{Trajs_Direct}
\end{figure}

\begin{figure}[t]
	\begin{center}
		\includegraphics[width=0.90\linewidth]{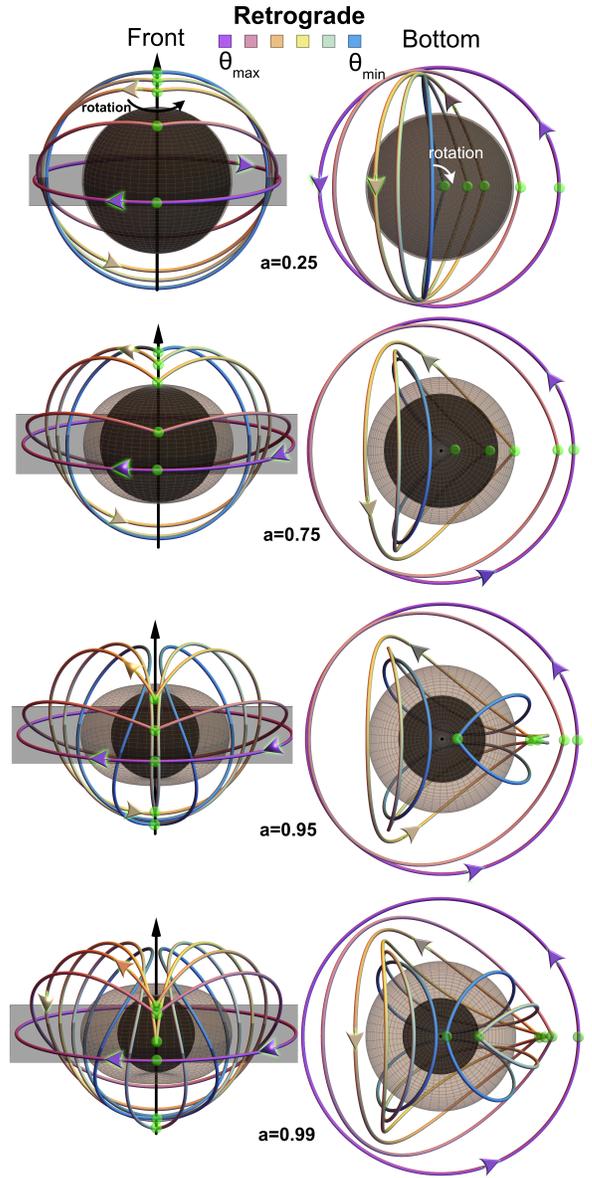}
	\end{center}
	\caption{ \emph{3-space projection of \underline{retrograde} trajectories.} Views are from front (left) and below (right). Arrows highlighted in green indicate direction of propagation, while the direction of black hole rotation is identified in the top panel of each column. The starting point of each trajectory is highlighted with a green sphere. Each row is associated with a particular value of black hole rotation parameter, $a$. Panels include the outer ergosphere (light gray) and the outer event horizon (dark gray).} 
\label{Trajs_Retrograde}
\end{figure}

Eqs. \ref{traj1} were then used, with the source/receiver positions of Fig. \ref{Initial_States}, to generate closed trajectories for representative send/receive points and values of rotation parameter $a$. These are shown in Figs. \ref{Trajs_Direct} and \ref{Trajs_Retrograde} for direct and retrograde orbits, respectively. Each trajectory is characterized by tangent vectors that are discontinuous  at the common location of the source and receiver. The associated misorientation of polarization constitutes a holonomy that can be engineered to take on a wide range of values.

%
%
%

%
%
%
%

\subsection{Evolution of Polarization for Spherical Trajectories}

For a specific trajectory, the evolution of polarization holonomy can be determined in two ways, both requiring only an initial polarization. The first method was detailed in section II D. It employs the null character of tangent vectors, polarization orthogonality with respect to propagation, and the Walker-Penrose constant $\mathbb{k}$, of Eq. \ref{k_Kerr}, to obtain an closed-form expression for the polarization at any point along a given trajectory. The second method, utilized here as a check on the analytical approach, numerically solves the parallel transport equation, Eq. \ref{fPX}. A subsequent gauge transformation is then implemented so that the resulting field has no temporal component at any point along the trajectory.  Both methods were used to produce evolving polarization states for a number of trajectories, and it was verified that the numerical results matched the analytical predictions. 

A given trajectory can support two orthogonal polarization vectors. Interestingly though, it is possible to construct initial polarization states such that only one vector exhibits a polarization holonomy. In particular, polarizations that are initially tangent to the local radial coordinate line never accumulate a polarization holonomy. Because of this, we need only construct the initial polarization to be orthogonal to such a radial polarization. The result can then be viewed as \emph{the} polarization holonomy of the trajectory.

\begin{figure}[t]
	\begin{center}
		\includegraphics[width=0.75\linewidth]{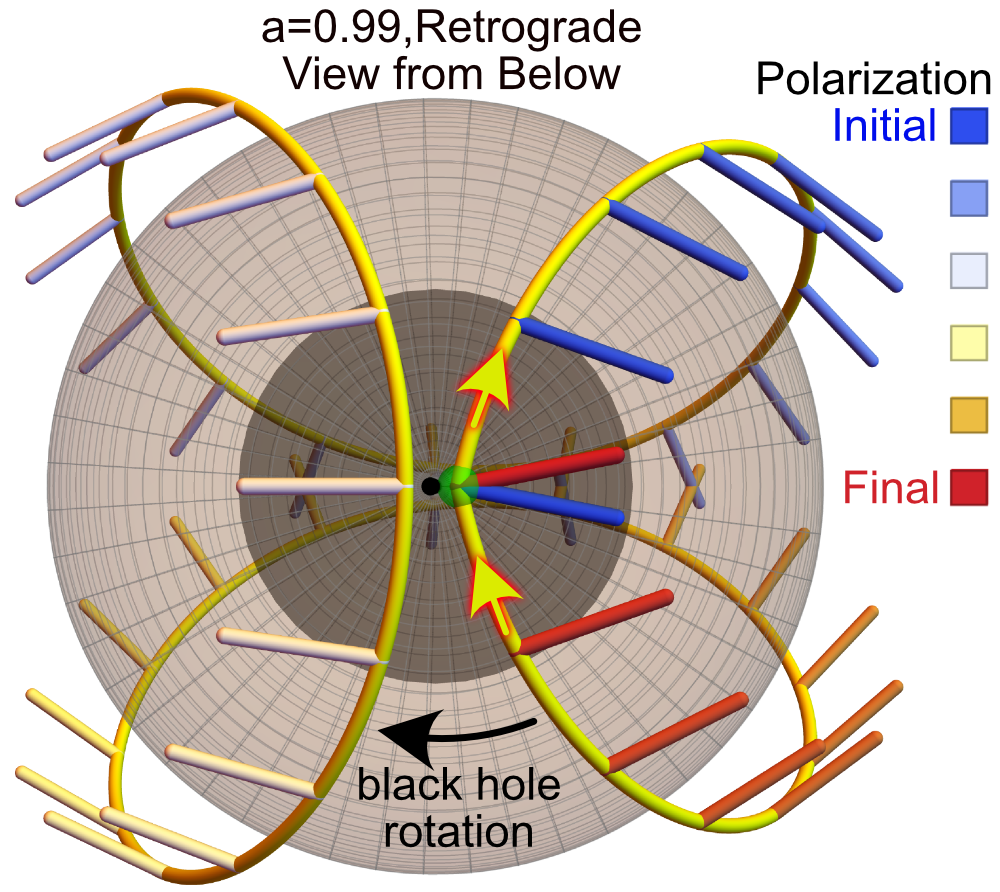}
	\end{center}
	\caption{ \emph{Evolution of Polarization}. The 3-space projection of a retrograde trajectory is shown in yellow for rotation parameter $a=0.99$ and start/end polar angle $\theta_0=177.3^\circ$ . Along this path, the polarization of light is plotted as colored sticks with color starting blue and ending red. The view is from below. Since the temporal component is engineered to always be zero, the angle between the red and blue sticks is the polarization holonomy. The figure includes the outer ergosphere (light gray) and the outer event horizon (dark gray).} 
	\label{a0p99_retro_polarization_below}
\end{figure}

A representative evolution of polarization is shown in Fig. \ref{a0p99_retro_polarization_below}. The angle between the initial (blue) and final (red) polarizations is the holonomy. 

%
%
%

\begin{figure}[t]
	\begin{center}
		\includegraphics[width=1.0\linewidth]{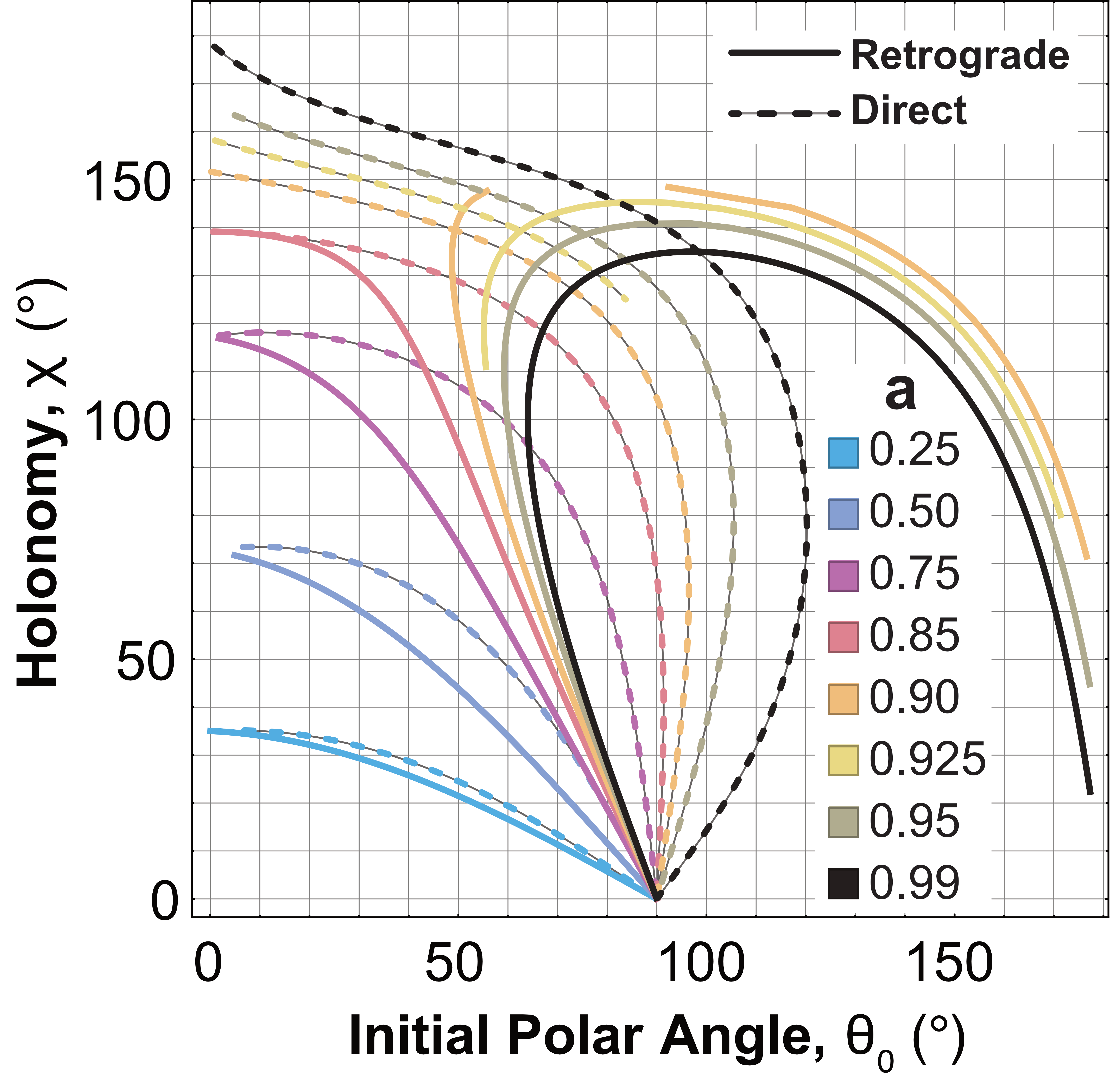}
	\end{center}
	\caption{ \emph{Polarization Holonomy}. The angle, $\chi$, between initial and final polarizations as a function of rotation parameter, $a$, and start/end polar angle, $\theta_0$. Results for retrograde trajectories are shown with solid curves while results for direct trajectories are dashed curves with a light gray outline to guide the eye.} 
	\label{Holonomy_Comparisons}
\end{figure}

Polarization holonomy can now be quantified for the entire class of closed trajectories. The procedure is to determine the angle of mismatch between between initial and final polarizations of each trajectory:
\begin{equation}\label{holonomy}
\chi = \cos^{-1}\left( {\bf f}_{\scriptstyle final} \cdot {\bf f}_{\scriptstyle init} \right).
\end{equation}
Results are plotted in Fig. \ref{Holonomy_Comparisons} for a spectrum of rotation parameters. In each case, the holonomy is calculated over the entire set of radii and polar angles for which a closed trajectory (of our identified class) is possible. The figure is intended to serve as a concise visual guide for understanding and engineering polarization holonomies.

%
%
%

\ml{Consistent with existing predictions, Fig. \ref{Holonomy_Comparisons} shows that there is no holonomy for equatorial transits\cite{FrolovShoom2011}. The classical Faraday analog is that propagation is parallel to the magnetic field so that there is no Coriolis acceleration. Likewise, there is no holonomy for singularities that do not rotate\cite{Ghosh_2016}. In a classical Faraday setting, this corresponds to a static magnetic field that is equal to zero. At the other extreme, the results indicate that direct circuits that start and end near the north pole result in polarization rotations of essentially $180^\circ$. Retrograde circuits that start and end near the south pole, on the other hand, generate a holonomy that is zero in the limit.}

\begin{figure}[t]
	\begin{center}
		\includegraphics[width=0.8\linewidth]{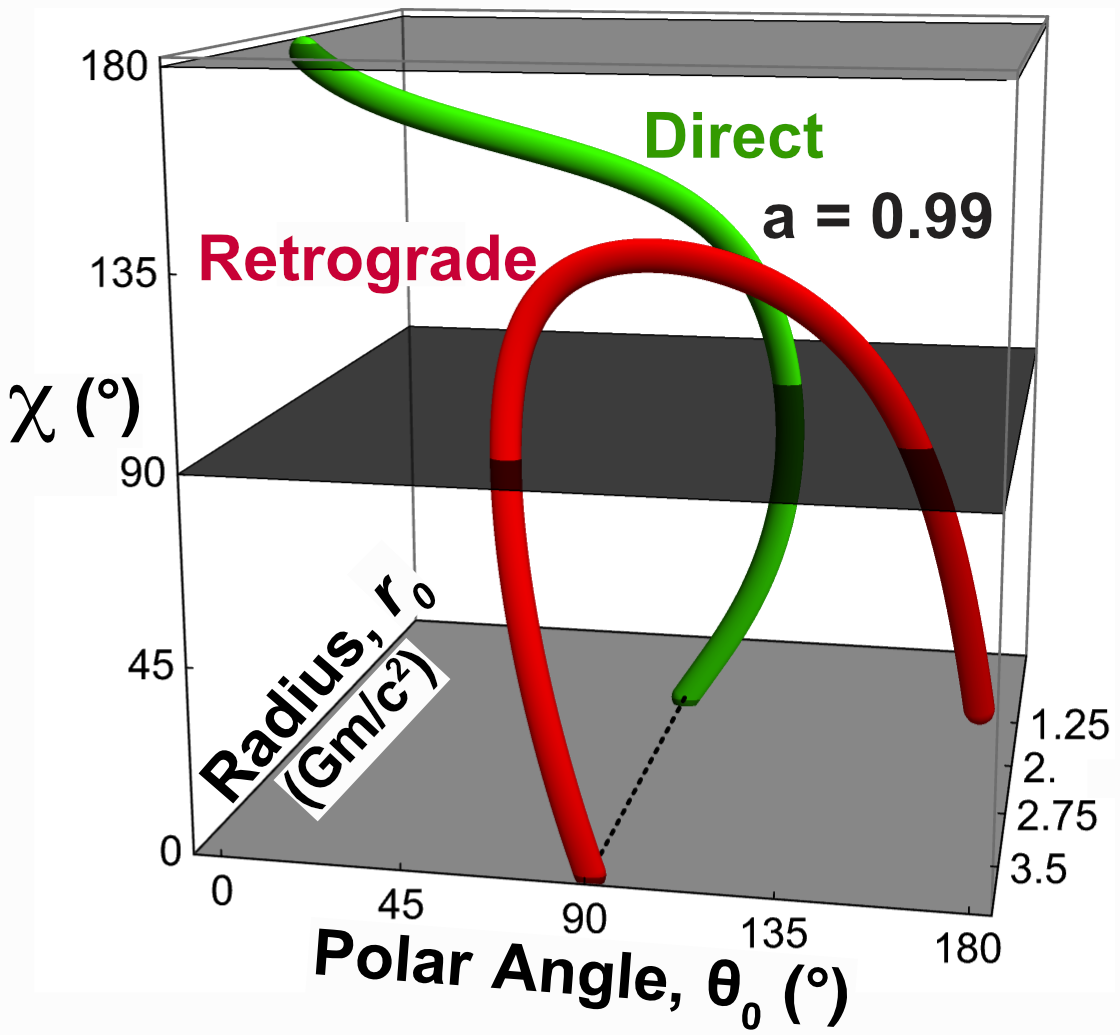}
	\end{center}
	\caption{ \emph{Polarization Holonomy as a Function of Polar Angle and Radius}. The holonomy angle, $\chi$, is plotted over the range of admissible values of both initial polar angle, $\theta_0$, and orbit radius, $r$, for $a = 0.99$. Retrograde data is shown in red and direct data in green. This shows that the holonomy data of Fig. \ref{Holonomy_Comparisons} is not multi-valued but, rather, is the result of projecting out all radial dependence.} 
	\label{holonomy_comparison_ap99_3D}
\end{figure}

\ml{While the holonomy appears to be multi-valued over a range of initial polar angles, the curves represent a range of radii. For instance, the solid black curve for retrograde motion around a rapidly rotating black hole, $a=0.99$, exhibits holonomies of both $\chi=0^\circ$ and $\chi=125^\circ$ for an initial polar angle of  $\theta_i = 90^\circ$. The former is associated with an equatorial transit, while the latter starts and ends at the equator but undergoes a complex circuit at a smaller radius. This can be seen in the lowest row of panels of Fig. \ref{Trajs_Retrograde}, but the three-dimensional plot Fig. \ref{holonomy_comparison_ap99_3D} makes this especially clear. There the holonomy is plotted as a function of both initial polar angle and radius for the direct and retrograde transits with $a=0.99$.}

%
%
%

\ml{Figs. \ref{Holonomy_Comparisons} and \ref{holonomy_comparison_ap99_3D} also demonstrate that direct and retrograde dynamics are markedly different, with both strong functions of black hole rotation rate. Beyond qualitative trends, Fig. \ref{Holonomy_Comparisons} shows that large projective polarization holonomies can be tailored and measured by sending and receiving light from a single stationary position. Such large holonomy angles should make experimental realization more tractable.  An experimental program could map out an entire spectrum of holonomy values to characterize spacetime in the vicinity of Kerr singularities. }

\section{Conclusions}
\ml{An important prediction of general relativity is that the plane of polarization of light will rotate in curved spacetime. Such Gravitational Faraday Rotation has been studied theoretically for decades, but it has yet to be experimentally measured. This is due to the difficulty in carrying out experiments in which light from a well-defined polarization source is subsequently measured at a distant receiver. Even if such measurements were to be carried out, the predicted rotation of the polarization plane is in on the order of milliradians with respect to a co-evolving Fermi-Walker reference frame\cite{Farooqui_2014}. The current proposal addresses these logistical and sensitivity issues. Light is emitted and received  from a single, stationary source, with the emission direction and polarization angle tailored so that rays travel in a nontrivial, closed circuit. The net changes in polarization rotation are on the order of radians, measured by projecting the final state onto the initial state.  A single experimental measurement of this sort would be an important confirmation of general relativity in strong-field settings.  Beyond this, though, a full range of direct and retrograde orbits can be used to interrogate Kerr spacetime, even the ergosphere. A mapping of the associated polarization rotation amounts to a characterization of spacetime local to rotating black holes.}

\ml{The proposed methodology is based on the holonomy exhibited when closed, three-space circuits of light have distinct outgoing and incoming polarizations.} This is not simply an artifact of the non-unique description of polarization associated with null-valued wave vectors.  Within the lowest-order geometric optics approximation in Kerr spacetime, a class of such circuits has been used to construct closed-form expressions for both the path and evolving polarization. Outgoing and incoming wave vectors are not aligned, a necessary but not sufficient condition for holonomy.  The results have been verified, for specific cases, by numerically solving the parallel transport equation for polarization. 

The Walker-Penrose Theorem was used to quantify polarization holonomy as a function of source/receiver radius and polar angle along with the black hole angular momentum per unit mass. For a given rotation parameter, $a$, the trajectories and associated holonomy can be smoothly changed by moving along well-defined initial position arcs. Both direct and retrograde paths were considered, and it was found that the latter are always associated with radii that are less than their direct counterparts. In addition, direct trajectories allow for holonomic geodesics that start and end within the ergosphere but emerge from it over an intermediate interval.

The approach of this work is in contrast to studies of Gravitational Faraday Rotation on open trajectories, in which a co-evolving frame must be used as a reference. In fact, the net rotation of polarization relative to a Fermi-Walker frame\cite{FrolovShoom2011} or a frame composed of principal null directions of the Weyl tensor\cite{Farooqui_2014} would be zero, verified in both cases. This is because the orientation of the frames themselves are discontinuous at the common location of source and receiver.

It is tempting to draw analogies between the polarization holonomies considered here and those associated with light propagating through inhomogeneous dielectrics\cite{Bliokh_2008, Iadecola2016, Shen2020, Chen2021}. The rotating singularities of the present work, though, correspond to three-dimensionally inhomogeneous, \emph{active} dielectrics\cite{MacKay_2010}, making it challenging to produce a laboratory surrogate\cite{Schuster_2024}.

There are a number of natural extensions to the present work. Perhaps the most straightforward of these would be to extend the results to include charged black holes. It would also be interesting to consider a higher-order geometric optics approximation\cite{Oancea_2020}, allowing the spin dependence of polarization holonomy to be determined. \ml{The large polarization rotation achievable lends promise to this.} For quantized radiation, pairs of entangled photons can be used to elucidate a black hole version of the Geometric Phase of Entanglement\cite{Sjoqvist_2000, Voitiv_2024}. Finally, it may be possible to identify holonomies associated with spatial modes of light as in, for instance, Laguerre-Gaussian, Hermite-Gaussian, and Bessel beams\cite{Galvez_2003, Lusk_2022}.

\section{Acknowledgments}
It is a pleasure to acknowledge useful discussions with Brendan McLane, Alex Tinguely, Andrew Voitiv, and Mark Siemens.

\appendix

\section{Equations of Motion}

Closed-form, spherical solutions can be constructed that satisfy the Equations of Motion, Eq \ref{EoM}. The process is tedious, but the ultimate result, Eqs. \ref{traj1}, is very practical and easy to use. Only the essential equations are provided below, since a detailed development is given elsewhere\cite{Gralla_2020, Wang_2022}.
\vskip 0.5 cm
\noindent Define $A$, $B$, and $C$.
\begin{align}
A &:= a^2 - \eta - \lambda^2 \nonumber \\
B &:= 2(\eta + (\lambda - a)^2) \nonumber \\
C &:= -a^2 \eta
\end{align}
Define $P$, $Q$, $w_\pm$ and $z$.
\begin{align}
P &:= -\frac{A^2}{12} - C \nonumber \\
Q &:= -\frac{A}{3} \left(  \left( \frac{A}{6} \right)^2 - C \right) - \frac{B^2}{8} \\
w_\pm &:= \left( \pm\sqrt{\frac{P^3}{27}+\frac{Q^2 }{4}}-\frac{Q}{2} \right)^{1/3}  \nonumber \\
z &:= \sqrt{\frac{w_+ + w_-}{2} - \frac{A}{6} } \nonumber 
\end{align}
Define key radii, $r_1$, $r_2$, $r_3$, and $r_4$.
\begin{align}
r_1 &:= -\sqrt{-\frac{A}{2}+\frac{B}{4
   z}-z^2} - z \nonumber \\
r_2 &:= \sqrt{-\frac{A}{2}+\frac{B}{4
   z}-z^2} - z \nonumber \\
r_3 &:=   -\sqrt{-\frac{A}{2}+\frac{B}{4
   z}-z^2} + z  \\
r_4 &:=   \sqrt{-\frac{A}{2}+\frac{B}{4
   z}-z^2} + z \nonumber \\
 r_\pm &:=   1 \pm \sqrt{1-a^2}\nonumber 
 \end{align}
Define $x_1$ and $k_1$.
\begin{align}
x_1 &:= \sqrt{\left( \frac{r - r_2}{r - r_1} \right) \left( \frac{r_3 - r_1}{r_3 - r_2}\right) } \nonumber \\
k_1 &:=\frac{\left(r_3-r_2\right)
   \left(r_4-r_1\right)}{\left
   (r_3-r_1\right)
   \left(r_4-r_2\right)}
\end{align}
Define $\Delta_\theta$ and $u_\pm$.
\begin{align}
\Delta_\theta &:= \frac{1}{2} \left( 1 - \frac{\eta + \lambda^2}{a^2} \right) \nonumber \\
u_\pm &:= \Delta_\theta  \pm \sqrt{\Delta_\theta^2 + \eta/a^2} 
\end{align}
Define $F_1$, $E_1$, $\Pi_1$, and $\Pi_\pm$. These are expressed in terms of Jacobi elliptic functions $E$, $F$, and $\Pi$.
\begin{align}
F_1 &:= \frac{2 F\left(\sin^{-1}\left(x_1\right) , k_1\right)}{\sqrt{\left(r_3-r_1\right) \left(r_4-r_2\right)}} \nonumber \\
 E_1 &:= \sqrt{\left(r_3 - r_1\right)  \left(r_4 - r_2\right) } E\left(\sin^{-1}\left(x_1\right) , k_1\right) \nonumber \\
 \Pi_1 &:= \frac{2 \Pi\left( \frac{r_3 - r_2}{r_3 - r_1}, \sin^{-1}\left(x_1\right) , k_1\right)}{\sqrt{\left(r_3-r_1\right) \left(r_4-r_2\right)}} \\
  \Pi_\pm &:= \frac{2 }{\sqrt{\left(r_3-r_1\right) \left(r_4-r_2\right)}} \frac{r_2 - r_1}{(r_\pm - r_1)(r_\pm - r_2)}  \nonumber \\ 
& \quad \times\Pi\left[ \left( \frac{r_\pm - r_1}{r_\pm - r_2}\right) \left( \frac{r_3 - r_2}{r_3 - r_1}\right) , \sin^{-1}\left(x_1\right) , k_1\right] \nonumber 
\end{align}
Define $\mathcal{I}_{0}$, $\mathcal{I}_{1}$, $\mathcal{I}_{2}$, and $\mathcal{I}_{\pm}$. 
\begin{align}
\mathcal{I}_{0} &:= F_1 \nonumber \\
\mathcal{I}_{1} &:= r_1 F_1 + (r_2 - r_1) \Pi_1 \nonumber \\
\mathcal{I}_{2} &:= \frac{\sqrt{\mathcal{R}}}{r - r_1} -  \frac{1}{2} (r_1 r_4 + r_2 r_3) F_1 - E_1   \\
\mathcal{I}_{\pm} &:=-\Pi_\pm -\frac{F_1}{r_\pm - r_1} \nonumber  
\end{align}
Define $\mathcal{G}_\theta$ and $\nu_\theta$.
\begin{align}
\mathcal{G}_\theta &:=\frac {-1} {\sqrt{-a^2 u_-}} F\left( \sin^{-1}\left( \frac{\cos \theta_{\scriptstyle init}}{\sqrt{u_+}} \right) ,\frac{u_+}{u_-}\right) \\
\nu_\theta &:= {\rm Sign(p^3_{\scriptstyle init})} \equiv {\rm Sign(p^\theta_{\scriptstyle init})} \nonumber 
\end{align}
Define $\mathcal{G}_\phi$ and $G_\phi$.
\begin{align}
\mathcal{G}_\phi &:=\frac {-1} {\sqrt{-a^2 u_-}} \Pi \left[ u_+,\sin^{-1}\left( \frac{\cos \theta_{\scriptstyle init}}{\sqrt{u_+}} \right) ,\frac{u_+}{u_-}\right] \nonumber \\
G_\phi &:=  - \nu_\theta \mathcal{G}_\phi \\
-&\frac {1}{\sqrt{-a^2 u_-}}  \Pi \left[ u_+, {\rm am} \left( \sqrt{-a^2 u_-} (s + \nu_\theta \mathcal{G}_\theta),  \frac{u_+}{u_-} \right) ,  \frac{u_+}{u_-} \right]   \nonumber 
\end{align}
Define functions $I_\pm (s)$ and $I(s)$.
\begin{align}
I_\pm (s) &:= \frac{s}{r - r_\pm} \nonumber \\
I(s) &:= \frac{4}{r_+ - r_-} \left[ r_+\left( r_+ - \frac{a \lambda}{2} \right) I_+(s)  \right] \nonumber \\ 
&\,\, - \frac{4}{r_+ - r_-} \left[ r_-\left( r_- - \frac{a \lambda}{2} \right) I_-(s)  \right]  \\
& \,\, + 2 r s + r^2 s + 4 s \nonumber 
\end{align}
Define function $G(s)$ in terms of Jacobi elliptic functions, $E$ and $F$, as well as the Jacobi amplitude function, ${\rm am}$.
%
\begin{align}
G(s) &:= \frac{-u_+}{\sqrt{-a^2 u_-}}  \left( \frac{u_-}{u_+} \right)  \nonumber \\
&\quad \times E\left[ \rm{am}\left( \sqrt{-a^2 u_-} (s + \nu_\theta\mathcal{G}_\theta), \frac{u_+}{u_-} \right), \frac{u_+}{u_-} \right]  \nonumber \\
& \,\,+ \frac{u_+}{\sqrt{-a^2 u_-}}  \left( \frac{u_-}{u_+} \right) \\
&\quad \times F\left[ \rm{am}\left( \sqrt{-a^2 u_-} (s + \nu_\theta\mathcal{G}_\theta), \frac{u_+}{u_-} \right), \frac{u_+}{u_-} \right]  \nonumber 
\end{align}

The expressions for spherical trajectories, r, can now be written out as a function of the Mino parameter, $s$:
\begin{align}\label{traj2}
\theta(s) &= \cos^{-1} \left[ -\nu_\theta \sqrt{u_+} \,\,{\rm sn} \left(  \sqrt{-a^2 u_-} (s + \nu_\theta \mathcal{G}_\theta ), \frac{u_+}{u_-} \right) \right] \nonumber \\
\phi(s) &= \lambda G_\phi  \nonumber\\
 + &\frac{2 a}{r_+ - r_-} \left[ \left( r_+ - \frac{a \lambda}{2} \right) \frac{s}{r - r_+} - \left( r_- - \frac{a \lambda}{2} \right) \frac{s}{r - r_-} \right] \nonumber \\
t(s) &= I(s) + a^2 G(s) . 
\end{align}


\end{document}